# New Indivisible Planetary Science Paradigm

J. Marvin Herndon
Transdyne Corporation
San Diego, CA 92131 USA

mherndon@san.rr.com

**Abstract:** I present here a new, indivisible planetary science paradigm, a wholly self-consistent vision of the nature of matter in the Solar System, and dynamics and energy sources of planets. Massive-core planets formed by condensing and raining-out from within giant gaseous protoplanets at high pressures and high temperatures. Earth's complete condensation included a ~300 Earth-mass gigantic gas/ice shell that compressed the rocky kernel to about 66% of Earth's present diameter. T-Tauri eruptions stripped the gases away from the inner planets and stripped a portion of Mercury's incompletely condensed protoplanet, and transported it to the region between Mars and Jupiter where it fused with in-falling oxidized condensate from the outer regions of the Solar System and formed the parent matter of ordinary chondrite meteorites, the main-Belt asteroids, and veneer for the inner planets, especially Mars. In response to decompression-driven planetary volume increases, cracks form to increase surface area and mountain ranges characterized by folding form to accommodate changes in curvature. The differences between the inner planets are primarily the consequence of different degrees of protoplanetary compression. The internal composition of Mercury is calculated by analogy with Earth. The rationale is provided for Mars potentially having a greater subsurface water reservoir capacity than before realized.

## Introduction

Images and data from orbiting spacecraft and landers have revealed new, important, unanticipated aspects of planets other than Earth. Understanding those observations, however, has posed a challenge for planetary investigators who generally are self-constrained within the framework of 'consensus favored' models they consider applicable to the formation of the terrestrial planets, in particular, the so-called 'standard model of solar system formation' dating from the 1960s, and a model of the internal composition of Earth which had its beginning circa 1940.

Interpretations of other planets are strongly colored by interpretations of our own, better-studied planet. In 1936, Lehmann discovered Earth's inner core [1]. At the time there was widespread belief that Earth resembled an ordinary chondrite meteorite. Within that circa 1940 understanding, the inner core's composition was thought to be iron metal in the process of crystallizing from the fluid iron alloy core [2]; the geomagnetic field was thought to be generated by convection-driven dynamo action in the fluid core, and; the rocky mantle surrounding the core was assumed to be of uniform composition with observed seismic discontinuities assumed to be caused by pressure-induced changes in crystal structure. Planetary investigators apply this interpretation of Earth to other planets, such as Mercury, but it is an incorrect interpretation.

I realized that discoveries made in the 1960s admitted a different possibility for the composition of the inner core, namely, fully crystallized nickel silicide [3]. That insight led me: (1) to evidence that Earth resembles, not an ordinary chondrite, but an enstatite chondrite; (2) to a fundamentally different interpretation of the composition of Earth's internal shells below a depth of 660 km and their state of oxidation; (3) to evidence of a new, powerful energy source and a different proposal for the generation-location of the geomagnetic field, and; (4) to a different understanding of Earth's formation and to new geodynamics that is the consequence. I have described the details and implications of this new, indivisible geoscience paradigm, called *Whole-Earth Decompression Dynamics (WEDD)*, in a number of scientific articles [4-13] and books [14-18]. 'Indivisible' in this instance means that the fundamental aspects of Earth are connected logically and causally, and can be deduced from our planet's early formation as a Jupiter-like gas giant.

The visionary evolutionist, Lynn Margulis, taught the importance of envisioning the Earth as a whole, rather than as unrelated segments spread among various scientific specialties [15]. In that spirit, and in the broader framework of the Solar System, I present here a new, indivisible planetary science paradigm, a wholly self-consistent vision of the nature of matter in the Solar System, and dynamics and energy sources of planets [5-8, 12, 13, 19-21], which differs profoundly from the half-century old, popular, but problematic paradigm. This is a new foundation from which much development is possible.

# 1. Problematic Planetary Science Paradigm

The first hypothesis about the origin of the Sun and the planets was advanced in the latter half of the 18th Century by Immanuel Kant and modified later by Pierre-Simon de Laplace. Early in the 20th Century, Laplace's nebula hypothesis was replaced with the Chamberlin-Moulton hypothesis which held that a passing star pulled matter from the Sun which condensed into large protoplanets and small planetesimals. Although the passing star idea fell out of favor, the nomenclature of protoplanets and planetesimals remained. Generally, concepts of planetary formation fall into one of two categories that involve either (1) condensation at high-pressures, hundreds to thousands of atm.; or (2) condensation at very low pressures.

Eucken [22] considered the thermodynamics of Earth condensing and raining-out within a giant gaseous protoplanet at pressures of 100-1000 atm. In the 1950s and early 1960s there was discussion of planetary formation at such pressures [23-25], but that largely changed with the 1963 publication by Cameron [26] of a model of solar system formation from a primordial gas of solar composition at low pressure, circa $10^{-4}$ atm.. Cameron's low pressure model became the basis for (1) condensation models that (wrongly) purported to produce minerals characteristic of ordinary chondrites as the equilibrium condensate from that medium [27, 28] and (2) planetary formation models based upon the Chamberlin-Moulton planetesimal hypothesis. The idea was that dust would condense from the gas at this very low pressure. Dust grains would collide with other grains, sticking together to become progressively larger grains, then pebbles, then rocks, then planetesimals and finally planets [29, 30].

Since the 1960s, the planetary science community almost unanimously concurred that Earth formed from primordial matter that condensed at a very low pressure, circa $10^{-4}$ atm. [27, 31]. The 'planetesimal hypothesis' was 'accepted' as the 'standard model of solar system formation'. However, as I discovered, there is an inherent flaw in that concept [5, 8, 32].

The inner planets all have massive cores, as known from their high relative densities. I was able to show by thermodynamic calculations that the condensate of primordial matter at those very low pressures would be oxidized, like the Orgueil C1/CI meteorite wherein virtually all elements are combined with oxygen. In such low pressure, low temperature condensate, there would be essentially no iron metal for the massive cores of the inner planets, a contradiction to the observation of massive-core planets.

The planetesimal hypothesis, *i.e.*, the 'standard model of solar system formation', is not only problematic from the standpoint of planetary bulk-density, but necessitates additional *ad hoc* hypotheses. One such necessary hypothesis is that of a radial Solar System temperature gradient during planetary formation, an assumed warm inner region delineated by a hypothetical 'frost line' between Mars and Jupiter; ice/gas condensation is assumed to occur only beyond that frost line. Another such necessary hypothesis is that of whole-planet melting, *i.e.*, the 'magma ocean', to account for core formation from essentially undifferentiated material. For other planetary systems with close-to-star gas giants, another such necessary hypothesis is that of 'planetary migration' where gas giants are assumed to form at Jupiter-distances from their star and then migrate inward.

## 2. Primary Mode of Planetary Formation

The above described popular version of planetary formation consists of an assemblage of assumption-based hypotheses that lack substantive connection with one another. That is not the case in the new, indivisible planetary science paradigm presented here: The highly-reduced state of primitive enstatite-chondrite matter is explained by high-pressure, high-temperature condensation from solar matter [5, 33] under circumstances similar to those derived by Eucken [22] for Earth raining out from within a giant gaseous protoplanet and the relative masses of inner parts of Earth, derived from seismic data, match corresponding, chemically-identified, relative masses of enstatite-chondrite-components (Table 1, Figure 1), observed by microscopic examination, indicating commonality of oxidation state and formation process.

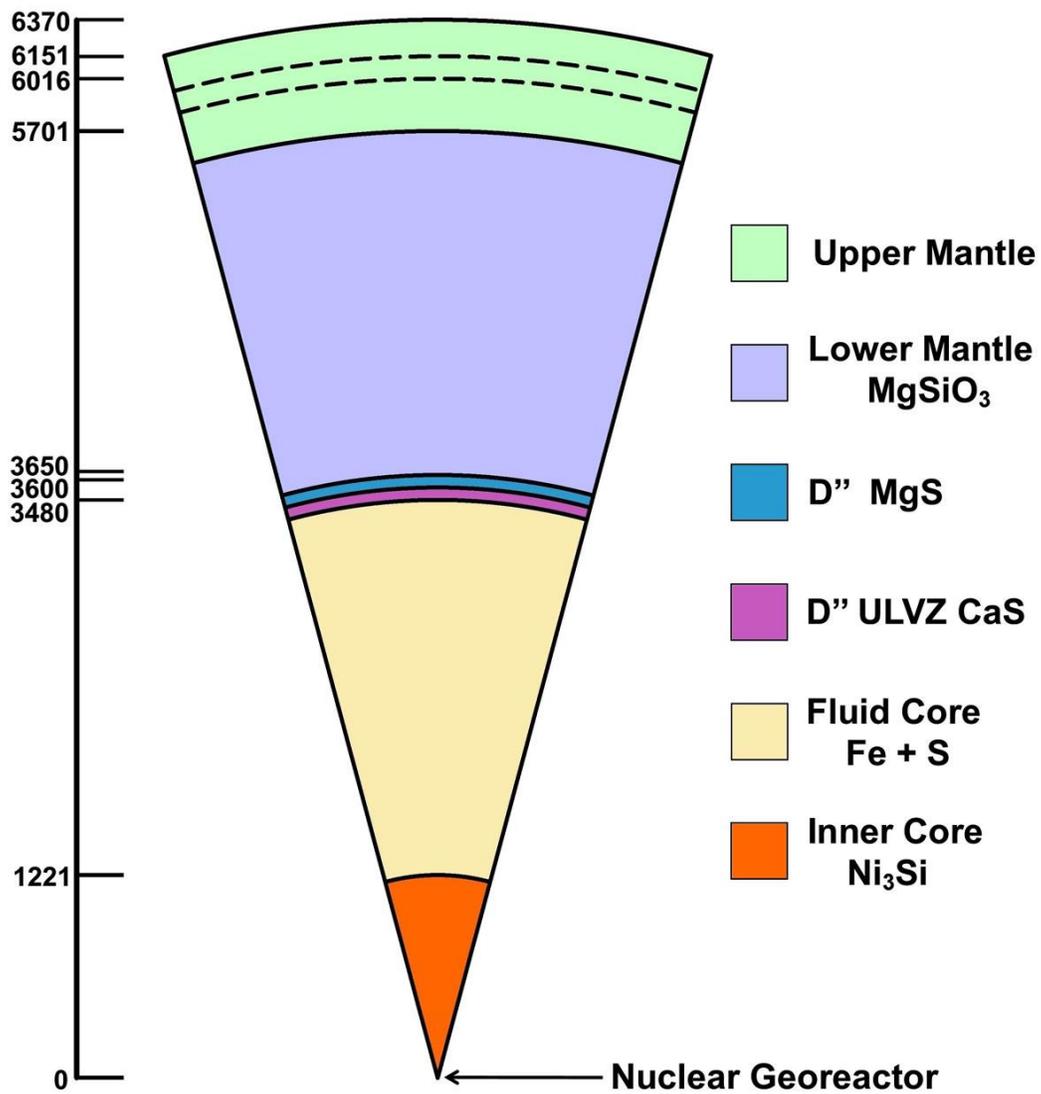

**Figure 1.** Chemical compositions of the major parts of the Earth, inferred from the Abee enstatite chondrite (see Table 1). The upper mantle, above the lower mantle, has seismically-resolved layers whose chemical compositions are not yet known. Radial distance scale in km.

**Table 1.** Fundamental mass ratio comparison between the endo-Earth (lower mantle plus core) and the Abee enstatite chondrite. Above a depth of 660 km seismic data indicate layers suggestive of veneer, possibly formed by the late addition of more oxidized chondrite and cometary matter, whose compositions cannot be specified with certainty at this time.

| Fundamental Earth Ratio | Earth Ratio Value | Abee Ratio Value |
|---|---|---|
| lower mantle mass to total core mass | 1.49 | 1.43 |
| inner core mass to total core mass | 0.052 | theoretical 0.052 if $Ni_3Si$ 0.057 if $Ni_2Si$ |
| inner core mass to lower mantle + total core mass | 0.021 | 0.021 |
| D″ mass to total core mass | 0.09*** | 0.11* |
| ULVZ** of D″ CaS mass to total core mass | 0.012**** | 0.012* |

\* = avg. of Abee, Indarch, and Adhi-Kot enstatite chondrites
D″ is the "seismically rough" region between the fluid core and lower mantle
\*\* ULVZ is the "Ultra Low Velocity Zone" of D″
\*\*\* calculated assuming average thickness of 200 km
\*\*\*\* calculated assuming average thickness of 28 km
data from [34-36]

Thermodynamic considerations led Eucken [22] to conceive of Earth formation from within a giant, gaseous protoplanet when molten iron rained out to form the core, followed by the condensation of the silicate-rock mantle. By similar, extended calculations I verified Eucken's results and deduced that oxygen-starved, highly-reduced matter characteristic of enstatite chondrites and by inference the Earth's interior, condensed at high temperatures and high pressures from primordial Solar System gas under circumstances that isolated the condensate from further reaction with the gas at low temperatures [5, 33].

In primordial matter of solar composition, there is a relationship between condensation pressure, condensation temperature, and the state of oxidation of the condensate. Ideally, when the partial pressure of a particular substance in the gas exceeds the vapor pressure of that condensed substance, the substance will begin to condense. In a gas of solar composition, the partial pressure of a substance is directly proportional to the total gas pressure, so at higher pressures substances condense at higher temperatures. The degree of oxidation of the condensate, on the other hand, is determined by the gas phase reaction

$$H_2 + \tfrac{1}{2}O_2 \leftrightarrow H_2O$$

which is a function of temperature but essentially independent of pressure. As I discovered, that reaction leads to an oxidized condensate at low temperatures and to a highly-reduced condensate at high temperatures, provided the condensate is isolated from further reaction with the gas [5, 33].

At pressures above about 1 atm. in a primordial atmosphere of solar composition, iron metal condenses as a liquid (Figure 2). That liquid can dissolve and sequester certain other elements, including significant hydrogen and a portion of oxygen-loving elements such as Ca, Mg, Si, and U. The composition and structure of the Earth's core (Figure 1) can be understood from the metallurgical behavior of an iron alloy of this composition initially with all of the core-elements fully dissolved at some high-temperature.

Elements with a high affinity for oxygen are generally incompatible in an iron alloy. So, when thermodynamically feasible those elements escaped from the liquid alloy. Calcium and magnesium formed CaS and MgS, respectively, which floated to the top of the core and formed the region referred to as D″. Silicon combined with nickel, presumably as $Ni_3Si$, and formed the inner core. The trace element uranium precipitated, presumably as US, and through one or more steps settled at the center of the Earth where it engaged in self-sustaining nuclear fission chain reactions [5, 20, 37-40].

The gaseous portion of primordial Solar System matter, as is the Sun's photosphere today, was about 300 times as massive as all of its rock-plus-metal forming elements. I posited Earth's complete condensation formed a gas-giant planet virtually identical in mass to Jupiter [4, 8, 15].

Giant gaseous planets of Jupiter size are observed in other planetary systems as close or closer to their star than Earth is to the Sun [41].

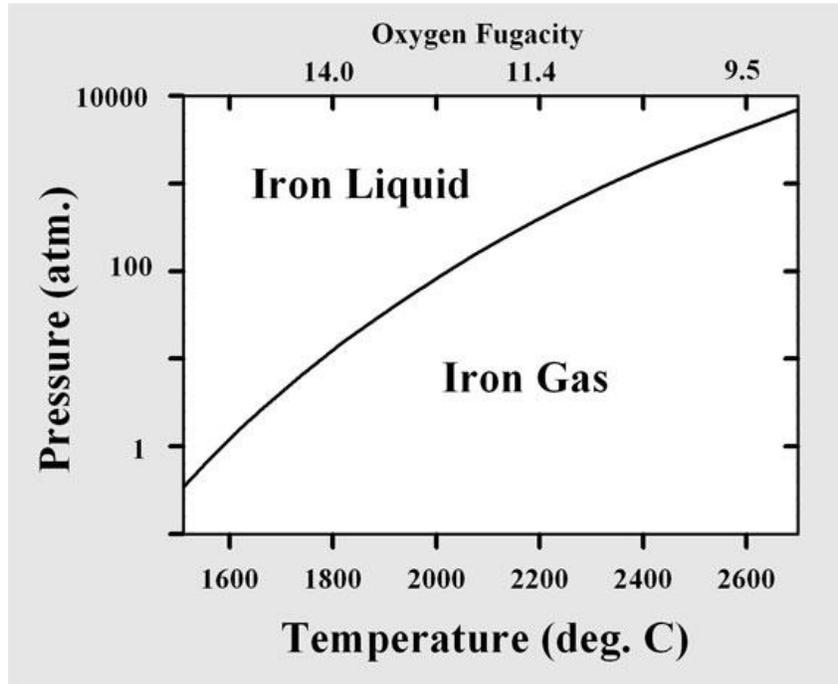

**Figure 2.** The curve in this figure shows the temperatures and total pressures in a cooling atmosphere of solar composition at which liquid iron will ideally begin to condense. The pressure-independent oxygen fugacity is shown on the upper abscissa.

Of the eight planets in the Solar System, the outer four (Jupiter, Saturn, Uranus, and Neptune) are gas-giants, whereas the inner four are rocky (Mercury, Venus, Earth, and Mars), without primary atmospheres. But the inner planets originated from giant gaseous protoplanets and their massive, primordial gases. How were the gases lost?

A brief period of violent activity, the T-Tauri phase, occurs during the early stages of star formation with grand eruptions and super-intense "solar-wind". The Hubble Space Telescope image of an erupting binary T-Tauri star is seen here in Figure 3. The white crescent shows the leading edge of the plume from a five-year earlier observation. The plume edge moved 130AU, a distance 130 times that from the Sun to Earth, in just five years. A T-Tauri outburst by our young Sun, I posit, stripped gas from the inner four planets. A rocky Earth, compressed by the weight of primordial gases, remained. Eventually Earth began to decompress driven primarily by the stored energy of protoplanetary compression. The consequences of Earth's formation in this manner provide rich new ways to interpret planetary data, especially when viewed in the broader context of Solar System processes responsible for the diversity of planet-forming matter.

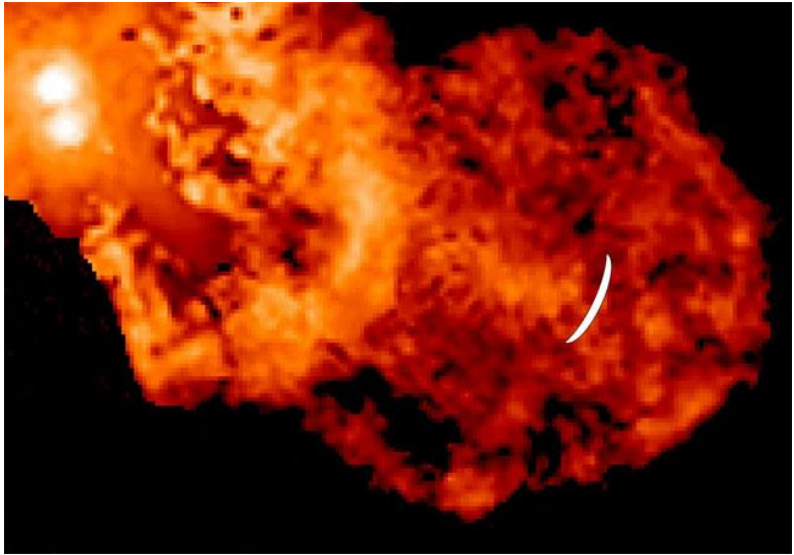

**Figure 3.** Hubble Space Telescope image of binary star XZ-Tauri in 2000 showing a T-Tauri phase outburst. The white crescent label shows the position of the leading edge of that plume in 1995, indicating a leading-edge advance of 130 A.U. in five years. T-Tauri eruptions are observed in newly formed stars. Such eruptions from our nearly-formed Sun, I submit, stripped the primordial gases from the inner four planets of our Solar System.

### 3. Matter of the Asteroid Belt, Mercury, and Ordinary Chondrites

The near-constancy in isotopic compositions of most of the elements of the Earth, the Moon, and the meteorites indicates formation from primordial matter of common origin [32]. Exceptions do occur and are important cosmochemical tracers, for example, oxygen and, in refractory inclusions of carbonaceous chondrites, magnesium, silicon, calcium, and titanium. Primordial elemental composition is yet evident to a great extent in the photosphere of the Sun and, for the less volatile, rock-forming elements, in chondrite meteorites, where many elements have not been separated from one another to within a factor of two. But there is complexity: rather than just one type of chondrite, there are three, with each type characterized by its own strikingly unique state of oxidation. Understanding the nature of the processes that yielded those three distinct types of matter from one common progenitor forms the basis for understanding much about planetary formation, their compositions, and the processes they manifest, including magnetic field production.

Only five major elements, iron (Fe), magnesium (Mg), silicon (Si), oxygen (O), and sulfur (S), comprise at least 95% of the mass of each chondrite and, by implication, each of the terrestrial planets. For decades, the abundances of major rock-forming elements ($E_i$) in chondrites have been expressed in the literature as atom ratios, usually relative to silicon ($E_i$/Si) and occasionally relative to magnesium ($E_i$/Mg). By expressing major-element abundances as molar (atom) ratios relative to iron ($E_i$/Fe), I discovered a fundamental relationship bearing on the genesis of chondrite matter, shown in Figure 4, which has implications on the nature of planetary processes in the Solar System [7]. Note in Figure 4 that the ordinary chondrite line intersects the other two. For this unique circumstance, each ordinary chondrite can be expressed as a linear combination of the compositions at the points of intersection. One intersection-component is a relatively undifferentiated carbonaceous-chondrite-like *primitive* component, with a state of oxidation like the Orgueil C1/CI chondrite, while the other is a partially differentiated enstatite-chondrite-like *planetary* component.

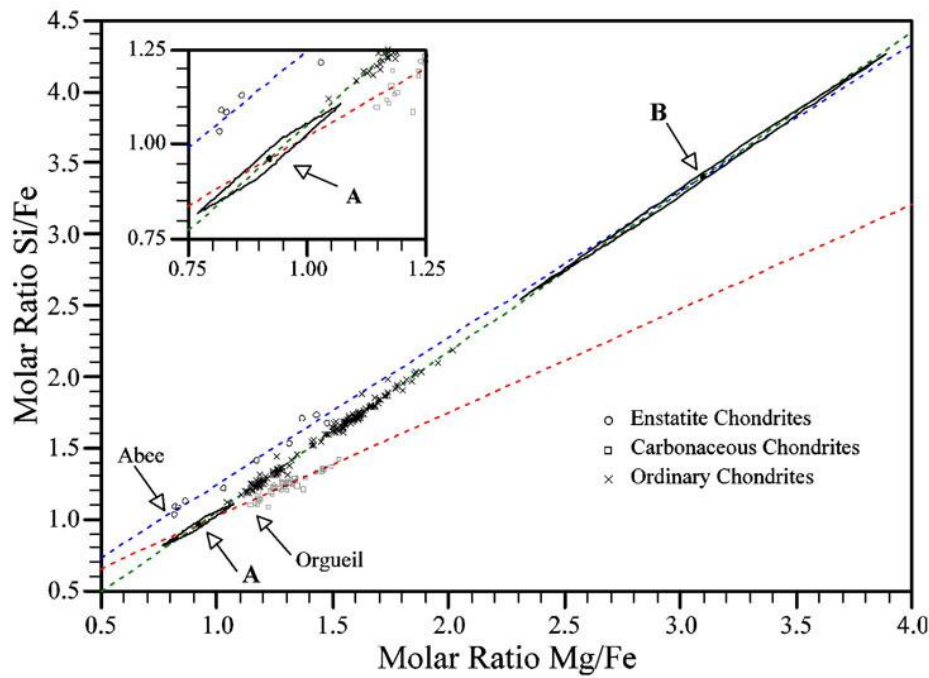

**Figure 4.** Molar (atom) ratios of Mg/Fe and Si/Fe from analytical data on 10 enstatite chondrites, 39 carbonaceous chondrites, and 157 ordinary chondrites. Data from [42-44]. Members of each chondrite class data set scatter about a unique, linear regression line. Upper line, enstatite chondrites; lower line carbonaceous chondrites, and; intersecting line, ordinary chondrites. The locations of the volatile-rich Orgueil carbonaceous chondrite and the volatile-rich Abee enstatite chondrite are indicated. Line intersections A and B are designated, respectively, *primitive* and *planetary* components. Error estimates of points A and B are indicated by solid-line parallelograms formed from the intersections of the standard errors of the respective linear regression lines. Inset shows in expanded detail the standard error parallelogram of point A.

Ordinary chondrites possess the common characteristic of being markedly depleted in refractory siderophile elements such as iridium and osmium. The degree of iridium and osmium depletion in each ordinary chondrite correlates with the relative proportion of its *planetary* component [7]. One can therefore conclude that the *planetary* component originated from a single large reservoir, characterized by a depletion in iridium and in osmium. From the inferred composition of the *planetary* component indicated in Figure 4, I suggested the partially-differentiated *planetary* component might be comprised of matter stripped from the protoplanet of incompletely-formed Mercury, presumably by the T-Tauri outbursts during thermonuclear ignition of the Sun. In the region between Mars and Jupiter, the ejected Mercury-component fused with in-falling Orgueil-like matter that had condensed at low pressures and low temperatures in the far reaches of the Solar System and/or in interstellar space. That fused combination become the parent matter of ordinary chondrites and asteroids of that region.

The molar (atom) Mg/Fe = 3.1 deduced for the *planetary* component indicates the stripping of Mercury's protoplanetary gases took place during the time when Mercury was only partially formed. The idea of heterogeneous protoplanetary differentiation/accretion is not new. Eucken [22] first suggested Earth's core formation as a consequence of successive condensation on the basis of relative volatility from a hot, gaseous protoplanet, with iron metal raining out at the center. The approximately seven-fold greater depletion within the *planetary* component of refractory siderophile elements (iridium and osmium) than other more volatile siderophile elements (nickel, cobalt, and gold) indicates that planetary-scale differentiation and/or accretion progressed in a heterogeneous manner. The first liquid iron to condense and rain-out preferentially scavenged the refractory siderophile elements from the hot gaseous protoplanet.

I estimated the original total mass of ordinary chondrite matter present in the Solar System as a function of the core mass of Mercury [7]. For a core mass equal to 75% of Mercury's present mass, the calculated original total ordinary chondrite mass amounts to $1.83 \times 10^{24}$ kg, about 5.5 times the mass of Mercury. That amount of mass is insufficient to have formed a planet as massive as the Earth, but may have contributed significantly to the formation of Mars, as well as adding a veneer to other planets, including Earth. Presently, only about 0.1% of that mass remains in the asteroid belt.

During the formation of the Solar System only three processes were primarily responsible for the diversity of matter in the Solar System and were directly responsible for planetary internal compositions and structures [5]. These are: (i) High-pressure, high-temperature condensation from primordial matter associated with planetary formation by raining-out from the interiors of giant-gaseous protoplanets; (ii) Low pressure, low temperature condensation from primordial matter in the remote reaches of the Solar System and/or in the interstellar medium associated with comets; and, (iii) Stripping of the primordial volatile components from the inner portion of the Solar System by super-intense T-Tauri phase outbursts during the thermonuclear ignition of the Sun. The internal composition of massive-core planets derives from (i) above, and leads to a simple commonality of highly-reduced internal planetary compositions. The outer portions of the terrestrial planets, however, appear in varying degree to be 'painted' by an additional veneer of more-oxidized matter derived from (ii) and (iii) above.

## 4. Inner Planets: Basis of Differences

Earth's surface is markedly different from that of the other inner planets in two pronounced ways: (1) About 41% of Earth's surface area is comprised of continental rock (sial) with the balance being ocean floor basalt (sima), and; (2) Like stitching on a baseball, a series of mid-ocean ridges encircles the Earth from which basalt extrudes, creeps across the ocean basins, and disappears into trenches. As disclosed in *Whole-Earth Decompression Dynamics (WEDD)*, these are consequences of Earth's early formation as a Jupiter-like gas giant with the rocky portion initially compressed to about 66% of present diameter by about 300 Earth-masses of primordial gases and ices [4].

Surface differences among the inner planets, I posit, are the consequence of circumstances that prevented the rocky kernels of other inner planets from being fully compressed by condensed gigantic gas/ice shells. As described above, stripping of Mercury's protoplanetary gases is inferred to have taken place during the time when Mercury was only partially formed [7]. One might speculate from relative density that the rocky kernel of Venus was fully formed, but the extent of its compression may differ from that of Earth due to the prevailing thermal environment and/or relative time of the Sun's T-Tauri outbursts. Eventually, the degree of compression experienced should be able to be estimated by understanding Venetian surface geology. Mars may be a special circumstance, having a relatively small, highly-reduced kernel surrounded by a relatively large shell of ordinary chondrite matter; additional information is needed to be more precise.

Earth's crust is markedly different from that of the other inner planets in harboring a geothermal gradient. Like Earth's two-component crust, the otherwise inexplicable geothermal gradient is understandable as a consequence of our planet's early formation as a Jupiter-like gas giant.

Since 1939, scientists have been measuring the heat flowing out of continental-rock [45, 46] and, since 1952, heat flowing out of ocean floor basalt [47]. Continental-rock contains much more of the long-lived radioactive nuclides than does ocean floor basalt. So, when the first heat flow measurements were reported on continental-rock, the heat was assumed to arise from radioactive decay. But later, ocean floor heat flow measurements, determined far from mid-ocean ridges, showed more heat flowing out of the ocean floor basalt than out of continental-rock measured away from heat-producing areas [48, 49]. This seemingly paradoxical result, I posit, arises from a previously unanticipated mode of heat transport that emplaces heat at the base of the crust. I call this mode of heat transport *Mantle Decompression Thermal Tsunami* [6].

Heat generated deep within the Earth may enhance mantle decompression by replacing the lost heat of protoplanetary compression. The resulting decompression, beginning within the mantle, will tend to propagate throughout the mantle, like a tsunami, until it reaches the impediment posed by the base of the crust. There, crustal rigidity opposes continued decompression; pressure builds and compresses matter at the mantle-crust-interface resulting in compression heating. This compression heating, I submit, is the source of heat that produces the geothermal gradient.

Earth's geothermal gradient serves as a barrier that limits the downward migration of water. The "geothermal gradient" is minimal or non-existent for terrestrial planets that lack the compression-stage characterized by an early, massive, fully condensed shell of primordial gases and ices. Mars appears to have lacked an early massive shell of compressive condensed gases. Without subsequent decompression of the Martian kernel, there is no basis to assume the existence of a "geothermal gradient"; there is no thermal barrier to the downward percolation of water. The absence of such a thermal barrier suggests that Mars may have a much greater subsurface water reservoir potential than previously realized.

In the popular, problematic planetary science paradigm, internal planetary heat is produced through the decay of long-lived radionuclides, the only non-hypothetical heat source, although for moons sometimes tidal friction is also included. In the new, indivisible planetary science paradigm described here, the following two important energy sources are added: (1) Stored energy of protoplanetary compression which, in the case of Earth, is the principle driving-energy for decompression and for heat emplacement at the base of the crust by *Mantle Decompression Thermal Tsunami*, and; (2) Planetocentric 'georeactor' nuclear fission energy.

During Earth's early formation as a Jupiter-like gas giant, the weight of ~300 Earth-masses of gas and ice compressed the rocky kernel to approximately 66% of present diameter. Because of rheology and crustal rigidity, the protoplanetary energy of compression was locked-in when the T-Tauri outbursts stripped away the massive gas/ice layer leaving behind a compressed kernel whose crust consisted entirely of continental rock (sial). Internal pressures began to build and eventually the crust began to crack.

To accommodate decompression-driven increases in volume in planetary volume, Earth's surface responds in two fundamentally different ways; by crack formation and by the formation of mountain chains characterized by folding.

Cracks form to increase the surface area required as a consequence of planetary-volume increases. *Primary* cracks are underlain by heat sources and are capable of basalt extrusion, for example, mid-ocean ridges; *secondary* cracks are those without heat sources, for example, submarine trenches, and which become the ultimate repositories for basalt extruded by *primary* cracks.

In addition to crack formation, decompression-increased planetary volume necessitates adjustments in surface curvature. Decompression-driven increases in volume result in a misfit of the continental rock surface formed earlier at a smaller Earth-diameter. This misfit results in 'excess' surface material confined within continent margins, which adjusts to the new surface curvature by buckling, breaking and falling over upon itself producing fold-mountain chains as illustrated in Figure 5 from [12].

Crack formation and the production of mountains characterized by folding, consequences of protoplanetary compression, are pronounced processes on Earth and may have some relevance to Venus. Planetocentric 'georeactor' nuclear fission energy, on the other hand, has relevance to virtually all planets and to some large moons.

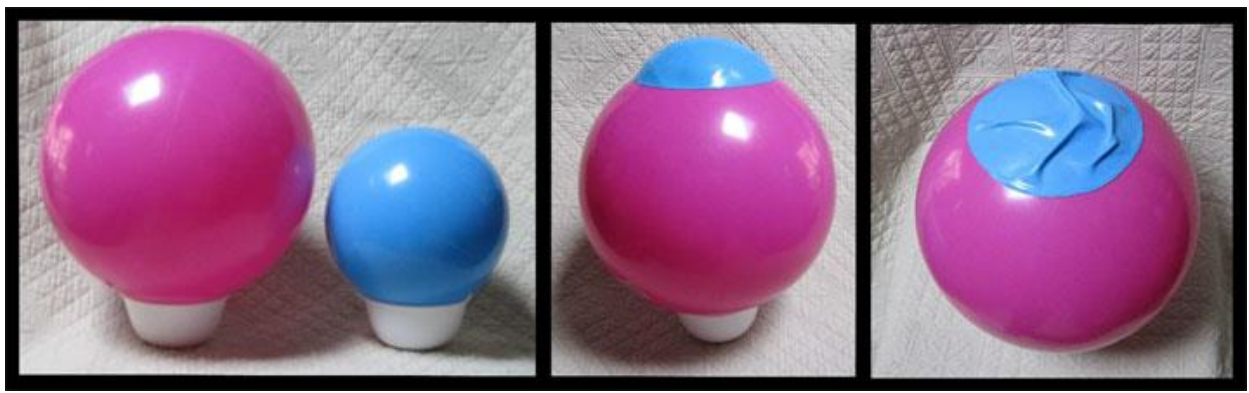

**Figure 5.** Demonstration illustrating the formation of fold-mountains as a consequence of Earth's early formation as a Jupiter-like gas giant. On the left, two balls representing the relative proportions of 'present' Earth (large), and 'ancient' Earth (small) before decompression. In the center, a spherical section, representing a continent, cut from 'ancient' Earth and placed on the 'present' Earth, showing: (1) the curvature of the 'ancient continent' does not match the curvature of the 'present' Earth and (2) the 'ancient continent' has 'extra' surface area confined within its fixed perimeter. On the right, tucks remove 'extra' surface area and illustrate the process of fold-mountain formation that is necessary for the 'ancient' continent to conform to the curvature of the 'present' Earth. Unlike the ball-material, rock is brittle so tucks in the Earth's crust would break and fall over upon themselves producing fold-mountains.

## 5. Evidence from Mercury's Surface

One of the most important Project MESSENGER discoveries were images from the spacecraft that revealed '… an unusual landform on Mercury, characterized by irregular shaped, shallow, rimless depressions, commonly in clusters and in association with high-reflectance material … and suggest that it indicates recent volatile-related activity' (Figure 6) and which have not been observed on any other rocky planet [50]. But the planetary investigators were unable to describe a scientific basis for the source of those volatiles or to suggest identification of the 'high-reflectance material'. I posited that during formation, condensing and raining-out as a liquid at high pressures and high temperatures from within a giant gaseous protoplanet, Mercury's iron alloy core dissolved copious amounts of hydrogen, one or more Mercury-volumes at STP. Hydrogen is quite soluble in liquid iron, but much less soluble in solid iron. I suggested that dissolved hydrogen from Mercury's core, released during core-solidification and escaping at the surface, produced hydrogen geysers that were responsible for forming those 'unusual landform on Mercury', sometimes referred to as pits or hollows, and for forming the associated 'high-reflectance material', bright spots, which I suggested is iron metal reduced from an exhaled iron compound, probably iron sulfide, by the escaping hydrogen [13].

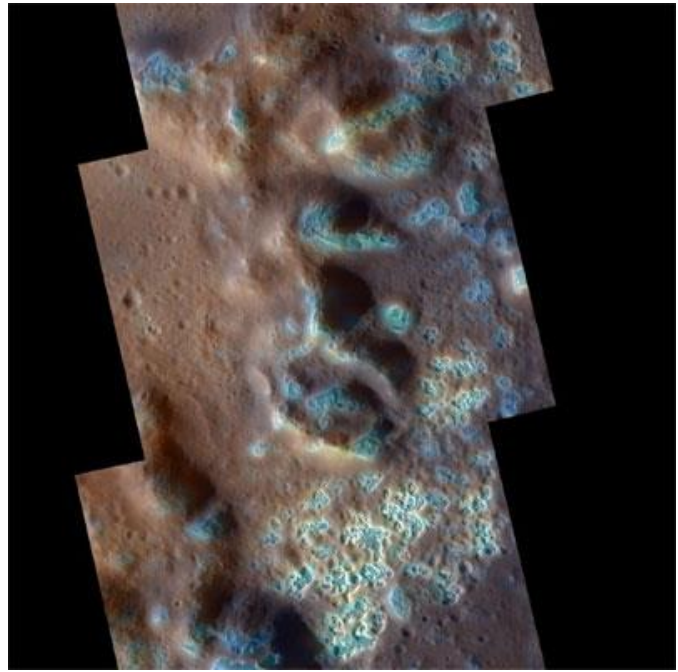

**Figure 6.** NASA MESSENGER image, taken with the Narrow Angle Camera, shows an area of hollows on the floor of Raditladi basin on Mercury. Surface hollows were first discovered on Mercury during MESSENGER's orbital mission and have not been seen on the Moon or on any other rocky planetary bodies. These bright, shallow depressions appear to have been formed by disgorged volatile material(s) from within the planet.

So, here is a test: Verifying that the 'high-reflectance material' is indeed metallic iron will not only provide strong evidence for Mercury's hydrogen geysers, but more generally will provide evidence that planetary interiors rained-out by condensing at high pressures and high temperatures within giant gaseous protoplanets. The high reflectance metallic iron can be distinguished by its low-nickel content from meteoritic metallic iron.

By analogy with Earth (Figure 1), the compositions of the interior parts of Mercury, calculated according to the mass ratio relationships presented in Table 1, are shown in Figure 7. Mercury's $MgSiO_3$ mantle mass is taken as the difference between planet mass and calculated core mass. Only 9 elements account for about 98% of the mass of a chondrite meteorite and the planet Mercury. Of the major and minor elements comprising Mercury's core, depicted in Figure 7, only aluminum and sodium, which have a high affinity for oxygen, are not represented. Presumably all aluminum and most, if not all, sodium occurs in Mercury's mantle/crust. Possibly a minor amount sodium might occur in Mercury's core as $NaCrS_2$ [51]. In the extreme case, if all of the trace element Cr formed $NaCrS_2$, a maximum of 18% of Mercury's sodium might occur as $NaCrS_2$.

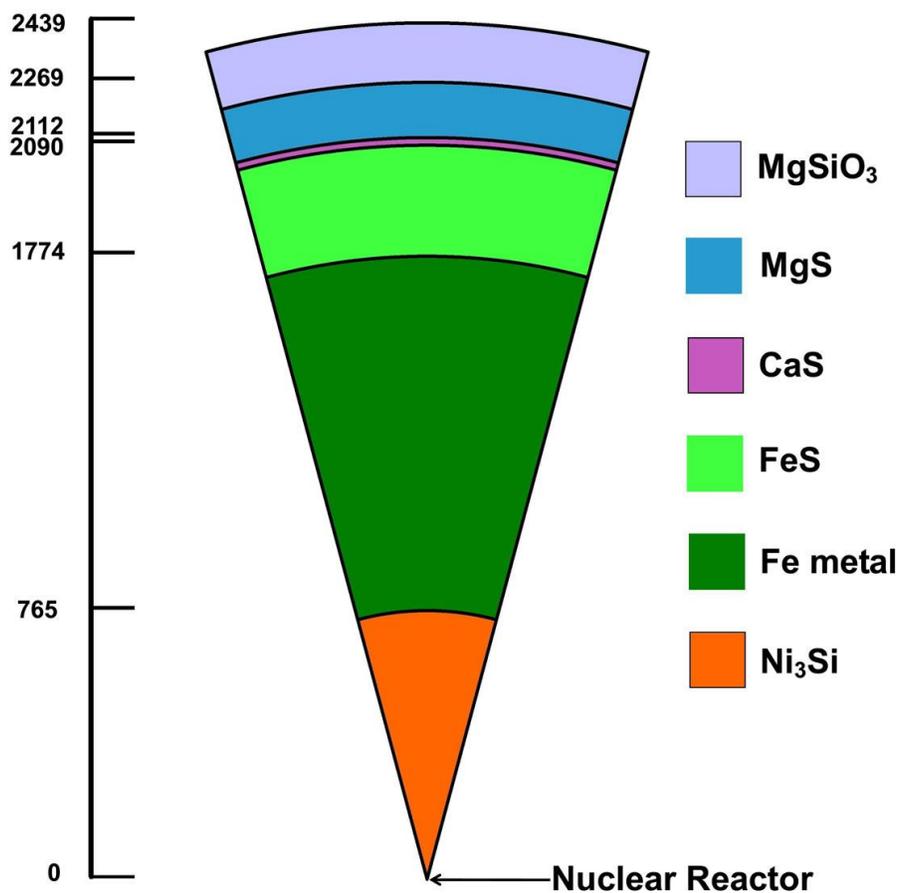

**Figure 7.** Internal structure of Mercury calculated from the mass ratio relationships of Earth shown in Table 1. Mercury's core is assumed to be fully solidified. The initial location of the planetocentric 'georeactor' is indicated. Radial distance scale in km.

As with Earth, the composition and structure of the Mercury's core (Figure 7) can be understood from the metallurgical behavior of an iron alloy initially with all of the core-elements fully dissolved at some high temperature. Upon cooling sufficiently, calcium and magnesium formed CaS and MgS, respectively, which floated to the top of the Mercurian core and formed the region analogous to Earth's D″. Silicon combined with nickel, presumably as $Ni_3Si$, and formed the inner core. The trace element uranium precipitated, presumably as US, and through one or more steps settled at Mercury's center where it inevitably engaged in self-sustaining nuclear fission chain reactions [8].

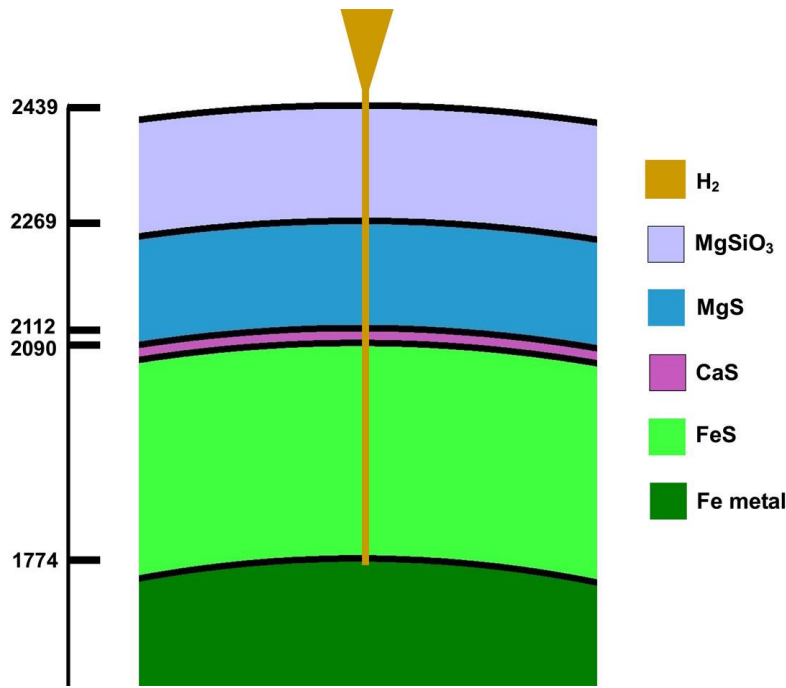

**Figure 8.** Schematic illustration of the source and path of hydrogen which is exhausted as hydrogen geysers and forms hollows (pits) on Mercury's surface. Radial distance scale in km.

One of the surprising early discoveries of the Project MESSENGER mission was abundant sulfur on Mercury's surface [52]. That observation is understandable as a consequence of hydrogen geysers. Figure 8 is a schematic representation of the path taken by exsolved hydrogen. Note the exiting hydrogen gas traverses regions of various sulfide compositions: iron sulfide (FeS), calcium sulfide (CaS), and magnesium sulfide (MgS). The exiting hydrogen, I submit, may scavenge sulfides from these layers and deposit them on Mercury's surface and perhaps may even emplace some in Mercury's exosphere.

Mercury is about 6% as massive as Earth. In the 1970's, this tiny planet's core, based upon heat-flow calculations, was thought to have solidified within the first billion years after formation [53]. But that was before my demonstration of the feasibility of planetocentric nuclear fission reactors [8, 21, 37] whose energy production considerably delayed solidification. Later, upon subsequent cooling, iron metal began to precipitate from Mercury's iron-sulfur alloy fluid core; the endpoint of core solidification is depicted in Figure 7. Core solidification with its concomitant release of dissolved hydrogen provides explanations for Mercurian surface phenomena.

## 6. Commonality of Nuclear Fission Heat and Magnetic Field Generation

Internally generated, currently active magnetic fields have been detected in six planets (Mercury, Earth, Jupiter, Saturn, Uranus and Neptune) and in one satellite (Jupiter's moon Ganymede). Magnetized surface areas of Mars and the Moon indicate the former existence of internally generated magnetic fields in those bodies. Furthermore, Jupiter, Saturn and Neptune radiate about twice as much energy as each receives from the Sun. Energy from nuclear fission chain reactions, part of the new, indivisible planetary science paradigm described here, provides logical and causally related explanations [8].

The condensate from within a giant gaseous protoplanet resembles an enstatite chondrite; thermodynamic condensation considerations are similar [5, 22, 33]. The interior of Earth, below 660 km, resembles an enstatite chondrite (Table 1). Thus, one may reasonably conclude that the Earth formed by raining out from within a giant gaseous protoplanet and that the interiors of other planets are similar to Earth's interior, which means their interiors are highly-reduced like the Abee enstatite chondrite. In the Abee meteorite, uranium occurs in the non-oxide part that corresponds to the Earth's core.

In cores of planets, density is a function of atomic number and atomic mass. Uranium, being the densest substance would tend ultimately to accumulate at the planets center. Applying Fermi's nuclear reactor theory, I demonstrated the feasibility of planetocentric nuclear fission reactors as energy sources for Jupiter, Saturn, and Neptune [19, 20] and for Earth as the energy source for the geomagnetic field [20, 37, 38]. Numerical simulations subsequently made at Oak Ridge National Laboratory verified those calculations and demonstrated that the georeactor could function over the entire age of the Earth as a fast neutron breeder reactor [39, 40]. Moreover, the calculations showed that helium would be produced in precisely the range of isotopic compositions observed exiting Earth.

The georeactor is a two-part assemblage, as illustrated in Figure 9, consisting of a fissioning nuclear sub-core surrounded by a sub-shell of radioactive waste products, presumably a liquid or slurry. The ~24 km diameter assemblage is too small to be presently resolved from seismic data.

Oceanic basalt helium data, however, provide strong evidence for the georeactor's existence [39, 54] and antineutrino measurements have not refuted that [55, 56]. To date, detectors at Kamioka, Japan and at Gran Sasso, Italy have detected antineutrinos coming from within the Earth. After years of data-taking, an upper limit on the georeactor nuclear fission contribution was determined to be either 26% (Kamioka, Japan) [56] or 15% (Gran Sasso, Italy) [55] of the total energy output of uranium and thorium, estimated from deep-Earth antineutrino measurements (Table 2). The actual total georeactor contribution may be somewhat greater, though, as some georeactor energy comes from natural decay as well as from nuclear fission.

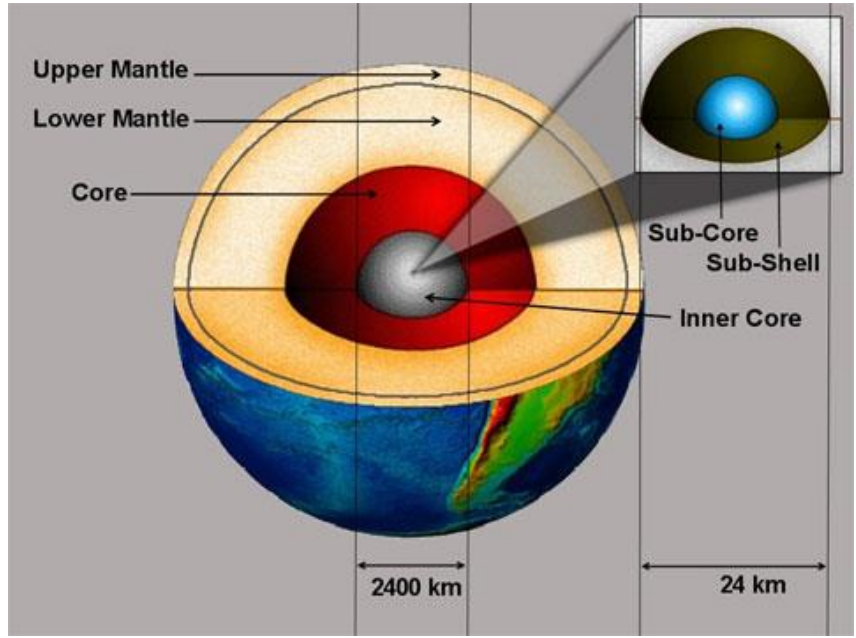

**Figure 9.** Earth's nuclear fission georeactor (inset) shown in relation to the major parts of Earth. The georeactor at the center is one ten-millionth the mass of Earth's fluid core. The georeactor sub-shell, I posit, is a liquid or a slurry and is situated between the nuclear-fission heat source and inner-core heat sink, assuring stable convection, necessary for sustained geomagnetic field production by convection-driven dynamo action in the georeactor sub-shell [8, 21, 38].

Before the Mariner 10 flybys in 1974 and 1975, in light of predictions of early core-solidification [53], there was essentially no expectation that Mercury possesses a currently generated magnetic field. That changed. The MESSENGER observations confirmed the existence of an actively generated, albeit very weak, global magnetic field centered close to the spin axis [57]. Efforts to explain Mercury's magnetic field generation within the problematic planetary science paradigm have proven to be challenging. This is why: Popular cosmochemical models fashioned on the idea that the internal composition of Mercury resembles an ordinary chondrite do not predict a substantial source of heat in Mercury's core. Without such a heat source, the core would solidify within about one billion years thus rendering core-convection impossible [53].

**Table 2.** Geoneutrino (antineutrino) determinations of radiogenic heat production [55, 56] shown for comparison with Earth's heat loss to space [58]. See original report for discussion and error estimates.

| Heat (terawatts) | Source |
| --- | --- |
| 44.2 TW | global heat loss to space |
| 20.0 TW | neutrino contribution from $^{238}$U, $^{232}$Th, and georeactor fission |
| 5.2 TW | georeactor KamLAND data |
| 3.0 TW | georeactor Borexino data |
| 4.0 TW | $^{40}$K theoretical |
| 20.2 TW | loss to space minus radiogenic |

In 1939, Elsasser first published his idea that the geomagnetic field is produced by convective motions in the Earth's fluid, electrically conducting core, interacting with rotation-produced Coriolis forces, creating a dynamo mechanism, a magnetic amplifier [59-61]. Elsasser's convection-driven dynamo mechanism seemed to explain so well the generation of the geomagnetic field that for decades geophysicists believed convection in the Earth's fluid core 'must' exist. Later, when it was discovered that many planets had internally generated magnetic fields, they were assumed, by analogy to Earth, to have convecting fluid iron alloy cores. But there is a problem, not with Elsasser's idea of a convection-driven dynamo, but with its location; as I discovered, convection is physically impossible in the Earth's fluid core and, presumably, as well in the cores of the various planets [8, 11].

Location of the source of the geomagnetic field in the georeactor sub-shell has implications for magnetic field reversals. The mass of the georeactor is only one ten-millionth the mass of the fluid core. High-intensity changing outbursts of solar wind, through the intermediary of the geomagnetic field, will induce electric currents into the georeactor, causing ohmic heating, which in extreme cases, might disrupt dynamo-convection and lead to a magnetic reversal. Massive trauma to the Earth might also disrupt sub-shell convection and lead to a magnetic reversal.

Why no Earth-core convection? The core is bottom-heavy, being approximately 23% denser at the bottom than at the top. The small decrease in density at the bottom due to thermal expansion is insufficient to overcome such a great density gradient. Moreover, for sustained convection the core-top must be maintained at a lower temperature than the core-bottom which is impossible because the Earth's core is wrapped in the mantle, a 2900 km thick thermally insulating blanket that has considerably lower thermal conductivity and heat capacity than the core.

In the popular problematic planetary science paradigm another problem is evident: There is no basis for the existence of a central heat source to drive the assumed planetary-core convection. But, in the new, indivisible planetary science paradigm described here, all of those problems are moot.

I have suggested that the geomagnetic field is produced by Elsasser's convection-driven dynamo operating within the georeactor's radioactive waste sub-shell [21]. Unlike the Earth's core, sustained convection appears quite feasible in the georeactor sub-shell. The top of the georeactor sub-shell is in contact with the inner core, a massive heat sink, which is in contact with the fluid core, another massive heat sink. Heat brought from the nuclear sub-core to the top of the georeactor sub-shell by convection is efficiently removed by these massive heat sinks thus maintaining the sub-shell adverse temperature gradient. Moreover, the sub-shell is not bottom heavy. Further, decay of neutron-rich radioactive waste in the sub-shell provides electrons that might provide the seed magnetic fields for amplification.

Among massive-core planets and large moons, there is a commonality of formation by condensing and raining-out of a gas of solar composition at high temperatures and high pressures, which leads to a commonality of internal compositions and highly-reduced states of oxidation, which in turn leads to a commonality of georeactor-like planetocentric nuclear fission reactors. In each case the central nuclear reactor is about one ten-millionth as massive as the planet's core and its operation does not depend upon the physical state of the core. That small mass means that major impacts could in principle offset the nuclear core from the planets center which, for example, might explain why Mercury's magnetic field is offset ~484 km north of center [57].

Venus currently has no internally generated magnetic field. Four potential explanations are: (1) Venus' rotation rate may be too slow; (2) Venus currently may be experiencing interrupted sub-shell convection such as might occur during a magnetic reversal; (3) Fuel breeding reactions at some point may have been insufficient for continued reactor operation, or; (4) Venus' 'georeactor' may have consumed all of its fissionable fuel. In light of helium evidence portending the eventual demise of Earth's georeactor [39], the fourth explanation seems most reasonable.

# 8. Summary

Massive-core planets formed by condensing and raining-out from within giant gaseous protoplanets at high pressures and high temperatures, accumulating heterogeneously on the basis of volatility with liquid core-formation preceding mantle-formation; the interior states of oxidation resemble that of the Abee enstatite chondrite. Core-composition was established during condensation based upon the relative solubilities of elements, including uranium, in liquid iron in equilibrium with an atmosphere of solar composition at high pressures and high temperatures. Uranium settled to the central region and formed planetary nuclear fission reactors, producing heat and planetary magnetic fields.

Earth's complete condensation included a ~300 Earth-mass gigantic gas/ice shell that compressed the rocky kernel to about 66% of Earth's present diameter. T-Tauri eruptions, associated with the thermonuclear ignition of the Sun, stripped the gases away from the Earth and the inner planets. The T-Tauri outbursts stripped a portion of Mercury's incompletely condensed protoplanet and transported it to the region between Mars and Jupiter where it fused with in-falling oxidized condensate from the outer regions of the Solar System and/or interstellar space, forming the parent matter of ordinary chondrite meteorites, the main-Belt asteroids, and veneer for the inner planets, especially Mars.

With its massive gas/ice shell removed, pressure began to build in the compressed rocky kernel of Earth and eventually the rigid crust began to crack. The major energy source for planetary decompression and for heat emplacement at the base of the crust is stored energy of protoplanetary compression is the stored energy of protoplanetary compression. In response to decompression-driven volume increases, cracks form to increase surface area and fold-mountain ranges form to accommodate changes in curvature.

One of the most profound mysteries of modern planetary science is this: As the terrestrial planets are more-or-less of common chondritic composition, how does one account for the marked differences in their surface dynamics? Differences among the inner planets are principally due to the degree of compression experienced. Planetocentric georeactor nuclear fission, responsible for magnetic field generation and concomitant heat production, is applicable to compressed and non-compressed planets and large moons.

The internal composition of Mercury is calculated based upon an analogy with the deep-Earth mass ratio relationships. The origin and implication of Mercurian hydrogen geysers is described. Besides Earth, only Venus appears to have sustained protoplanetary compression; the degree of which might eventually be estimated from understanding Venetian surface geology. A basis is provided for understanding that Mars essentially lacks a 'geothermal gradient' which implies potentially greater subsurface water reservoir capacity than previously expected.

# References

1. Lehmann, I., P'. *Publ. Int. Geod. Geophys. Union, Assoc. Seismol., Ser. A, Trav. Sci.*, 1936, **14**, 87-115.

2. Birch, F., The transformation of iron at high pressures, and the problem of the earth's magnetism. *Am. J. Sci.*, 1940, **238**, 192-211.

3. Herndon, J. M., The nickel silicide inner core of the Earth. *Proc. R. Soc. Lond*, 1979, **A368**, 495-500.

4. Herndon, J. M., Whole-Earth decompression dynamics. *Curr. Sci.*, 2005, **89**(10), 1937-1941.

5. Herndon, J. M., Solar System processes underlying planetary formation, geodynamics, and the georeactor. *Earth, Moon, and Planets*, 2006, **99**(1), 53-99.

6. Herndon, J. M., Energy for geodynamics: Mantle decompression thermal tsunami. *Curr. Sci.*, 2006, **90**, 1605-1606.

7. Herndon, J. M., Discovery of fundamental mass ratio relationships of whole-rock chondritic major elements: Implications on ordinary chondrite formation and on planet Mercury's composition. *Curr. Sci.*, 2007, **93**(3), 394-398.

8. Herndon, J. M., Nature of planetary matter and magnetic field generation in the solar system. *Curr. Sci.*, 2009, **96**, 1033-1039.

9. Herndon, J. M., Impact of recent discoveries on petroleum and natural gas exploration: Emphasis on India. *Curr. Sci.*, 2010, **98**, 772-779.

10. Herndon, J. M., Potentially significant source of error in magnetic paleolatitude determinations. *Curr. Sci.*, 2011, **101**(3), 277-278.

11. Herndon, J. M., Geodynamic Basis of Heat Transport in the Earth. *Curr. Sci.*, 2011, **101**, 1440-1450.

12. Herndon, J. M., Origin of mountains and primary initiation of submarine canyons: the consequences of Earth's early formation as a Jupiter-like gas giant. *Curr. Sci.*, 2012. **102**(10), 1370-1372.

13. Herndon, J. M., Hydrogen geysers: Explanation for observed evidence of geologically recent volatile-related activity on Mercury's surface. *Curr. Sci.*, 2012. **103**(4), 361-361.

14.	Herndon, J. M., *Maverick's Earth and Universe*, 2008, Vancouver: Trafford Publishing. ISBN 978-1-4251-4132-5.

15.	Herndon, J. M., *Indivisible Earth: Consequences of Earth's Early Formation as a Jupiter-Like Gas Giant*, L. Margulis, Editor 2012, Thinker Media, Inc.

16.	Herndon, J. M., Beyond Plate Tectonics: Consequence of Earth's Early Formation as a Jupiter-Like Gas Giant, 2012, Thinker Media, Inc.

17.	Herndon, J. M., *Origin of the Geomagnetic Field: Consequence of Earth's Early Formation as a Jupiter-Like Gas Giant*, 2012, Thinker Media, Inc.

18.	Herndon, J. M., *What Meteorites Tell Us About Earth*, 2012, Thinker Media, Inc.

19.	Herndon, J. M., Nuclear fission reactors as energy sources for the giant outer planets. *Naturwissenschaften*, 1992, **79**. 7-14.

20.	Herndon, J. M., Planetary and protostellar nuclear fission: Implications for planetary change, stellar ignition and dark matter. *Proc. R. Soc. Lond*, 1994, **A455**, 453-461.

21.	Herndon, J. M., Nuclear georeactor generation of the earth's geomagnetic field. *Curr. Sci.*, 2007, **93**(11), 1485-1487.

22.	Eucken, A., Physikalisch-chemische Betrachtungen ueber die frueheste Entwicklungsgeschichte der Erde. *Nachr. Akad. Wiss. Goettingen, Math.-Kl.*, 1944, 1-25.

23.	Bainbridge, J., Gas imperfections and physical conditions in gaseous spheres of lunar mass. *Astrophys. J.*, 1962, **136**, 202-210.

24.	Kuiper, G. P., On the origin of the Solar System. *Proc. Nat. Acad. Sci. USA*, 1951, **37**, 1-14.

25.	Kuiper, G. P., On the evolution of the protoplanets. *Proc. Nat. Acad. Sci. USA*, 1951, **37**, 383-393.

26.	Cameron, A. G. W., Formation of the solar nebula. *Icarus*, 1963, **1**, 339-342.

27.	Grossman, L., Condensation in the primitive solar nebula. *Geochim. Cosmochim. Acta*, 1972, **36**, 597-619.

28.	Larimer, J. W. and Anders, E., Chemical fractionations in meteorites-III. Major element fractionations in chondrites. *Geochim. Cosmochim. Acta*, 1970, **34**, 367-387.

29.	Goldrich, P. and Ward, W. R., The formation of planetesimals. *Astrophys J.*, 1973, **183**(3),1051-1061.

30. Wetherill, G. W., Formation of the terrestrial planets. *Ann. Rev. Astron. Astrophys.*, 1980, **18**: p. 77-113.

31. Larimer, J. W., Chemistry of the solar nebula. *Space Sci. Rev.*, 1973, **15**(1), 103-119.

32. Herndon, J. M., Reevaporation of condensed matter during the formation of the solar system. *Proc. R. Soc. Lond*, 1978, **A363**, 283-288.

33. Herndon, J. M. and Suess, H. E., Can enstatite meteorites form from a nebula of solar composition? *Geochim. Cosmochim. Acta*, 1976, **40**, 395-399.

34. Dziewonski, A. M. and Anderson, D. A., Preliminary reference Earth model. *Phys. Earth Planet. Inter.*, 1981, **25**, 297-356.

35. Keil, K., Mineralogical and chemical relationships among enstatite chondrites. *J. Geophys. Res.*, 1968, **73**(22), 6945-6976.

36. Kennet, B. L. N., Engdahl, E. R. and Buland, R., Constraints on seismic velocities in the earth from travel times. *Geophys. J. Int.*, 1995, **122**, 108-124.

37. Herndon, J. M., Feasibility of a nuclear fission reactor at the center of the Earth as the energy source for the geomagnetic field. *J. Geomag. Geoelectr.*, 1993, **45**, 423-437.

38. Herndon, J. M., Sub-structure of the inner core of the earth. *Proc. Nat. Acad. Sci. USA*, 1996, **93**, 646-648.

39. Herndon, J. M., Nuclear georeactor origin of oceanic basalt $^3$He/$^4$He, evidence, and implications. *Proc. Nat. Acad. Sci. USA*, 2003, **100**(6), 3047-3050.

40. Hollenbach, D. F. and Herndon, J. M., Deep-earth reactor: nuclear fission, helium, and the geomagnetic field. *Proc. Nat. Acad. Sci. USA*, 2001, **98**(20), 11085-11090.

41. Seager, S. and Deming, D., Exoplanet Atmospheres. *Ann. Rev. Astron. Astrophys.*, 2010, **48**, 631-672.

42. Baedecker, P.A. and Wasson, J. T., Elemental fractionations among enstatite chondrites. *Geochim. Cosmochim. Acta*, 1975, **39**, 735-765.

43. Jarosewich, E., Chemical analyses of meteorites: A compilation of stony and iron meteorite analyses. *Meteoritics*, 1990, **25**, 323-337.

44. Wiik, H. B., On regular discontinuities in the composition of meteorites. *Commentationes Physico-Mathematicae*, 1969, **34**, 135-145.

45. Benfield, A. F., Terrestrial heat flow in Great Britain. *Proc. R. Soc. Lond*, 1939, **A173**, 428-450.

46. Bullard, E. C., Heat flow in South Africa. *Proc. R. Soc. Lond*, 1939, **A173**, 474-502.

47. Revelle, R. and Maxwell, A. E., Heat flow through the floor of the eastern North Pacific Ocean. Nature, 1952, **170**, 199-200.

48. Blackwell, D. D., The thermal structure of continental crust, in *The Structure and Physical Properties of the Earth's Crust, Geophysical Monograph 14*, J. G. Heacock, Editor 1971, American Geophysical Union: Washington, DC., p. 169-184.

49. Stein, C. and Stein, S., A model for the global variation in oceanic depth and heat flow with lithospheric age. *Nature*, 1992, **359**, 123-129.

50. Blewett, D. T., et al., Hollows on Mercury: MESSENGER Evidence for Geologically Recent Volatile-Related Activity. *Sci.*, 2011, **333**, 1859-1859.

51. Okada, A. and Keil, K., Caswellsilverite, $NaCrS_2$: A new mineral in the Norton County enstatite achondrite. *Am. Min.*, 1982, **67**, 132-136.

52. Nittler, L. R., et al., The major element composition of Mercury's surface from MESSENGER X-ray spectrometry. *Sci.*, 2011, **333**, 1847-1850.

53. Solomon, S. C., Some aspects of core formation in Mercury. *Icarus*, 1976, **28**(4), 509-521.

54. Rao, K. R., Nuclear reactor at the core of the Earth! - A solution to the riddles of relative abundances of helium isotopes and geomagnetic field variability. *Curr. Sci.*, 2002, **82**(2), 126-127.

55. Bellini, G., et al., Observation of geo-neutrinos. *Phys. Lett.*, 2010, **B687**, 299-304.

56. Gando, A., et al., Partial radiogenic heat model for Earth revealed by geoneutrino measurements. *Nature Geosci.*, 2011, **4**, 647-651.

57. Anderson, B. J., et al., The global magnetic field of Mercury from MESSENGER orbital observations. *Sci.*, 2011., **333**, 1859-1862.

58. Pollack, H. N., Hurter, S. J., Johnson, J. R., Heat flow from the Earth's interior: Analysis of the global data set. *Rev. Geophys.*, 1993, **31**(3), 267-280.

59. Elsasser, W. M., On the origin of the Earth's magnetic field. *Phys. Rev.*, 1939. **55**, 489-498.

60. Elsasser, W. M., Induction effects in terrestrial magnetism. Phys. Rev., 1946, **69**, 106-116.

61. Elsasser, W. M., The Earth's interior and geomagnetism. *Revs. Mod. Phys.*, 1950, **22**, 1-35.